# On the role of Lifshitz invariants in liquid crystals

A. Sparavigna

**Dipartimento di Fisica**
**Politecnico di Torino**
**Corso Duca degli Abruzzi 24, Torino, Italy**

**Abstract**
The interaction between an external action and the order parameter, via a dependence described by a so-called Lifshitz invariant, is very important to determine the final configuration of the liquid crystal cells. The external action can be an electric field applied to the bulk or the confinement due to free surfaces or cell walls. The Lifshitz invariant includes the order parameter in the form of an elastic strain. This coupling between elastic strains and fields, inserted in a Landau-Ginzburg formalism, is well known and give rise to striction effects causing undulations in the director configuration. We want to discuss here the role of Lifshitz coupling terms, following an approach similar to that introduced by Dzyaloshinskii for magnetic materials. Case studies on nematics in planar and cylindrical cells are also proposed.

**1. Introduction**
The contributions to the free-energy density of terms in the derivatives of order parameter are of great importance and recognised to be fundamental in governing the appearance of spatially modulated structures in magnetic materials and of periodic patterns in liquid crystals. It is possible to identify the same structure in the free energy, when it is represented by a Landau-Lifshitz phenomenological theory of phase transitions: this structure has the form of an invariant term, so-called Lifshitz invariant, which is linear with respect to the gradient of order parameter. As shown by Lifshitz [1], a system near its phase transition point may be unstable with respect to distortions of the appropriate order parameter. This instability may develop if the irreducible representation allows a quadratic antisymmetric combination, linear both in the order parameter components and in their gradients.
Phases with large-scale space fluctuations of the order parameter were discovered experimentally in the 1960's [2]. Using the approach proposed by Lifshitz, Dzyaloshinskii [3] showed that these configurations are associated with the development of instabilities and found the corresponding approximate solutions of the phase equations. Now, the family of experimentally observed modulated states has growth both in magnetic and liquid crystal systems [4-7].
One of the aims of this paper is to find and enhance all the features common to magnetic and liquid crystal materials. For this reason, the first part of the paper starts with a brief remark on the use of Lifshitz invariant in magnetic systems. Then we discuss the invariants in liquid crystals; in the last part of the paper, we will see some case studies. In particular, we propose a detailed discussion of the hybrid cell with phase diagrams including the flexoelectric effect. The same we do for the cylindrical confinement of nematics.

## 2. The Dzyaloshinskii-Moriya coupling

Some magnetic structures are characterised by a modulation of the spin arrangements over periods, which are long compared to the size of the lattice cell and usually not commensurate with it. The existence of such magnetic structures can be due to competition between exchange interactions or to relativistic effects like spin-orbit coupling. Relativistic interactions were first considered by Dzyaloshinskii [3] and received a microscopic description by Moriya [8]. The Dzyaloshinskii-Moriya (DM) interaction can be written as a cross product of three vectors $F_{DM} = \vec{D} \cdot (\vec{S}_i \times \vec{S}_j)$ where $\vec{D}$ is the DM-vector and $\vec{S}_i, \vec{S}_j$ are spin vectors. The bond symmetry determines the direction of the DM-vector whereas the strength of the spin-orbit coupling gives its intensity [8,9].

The macroscopic manifestation of the antisymmetric DM couplings takes place in non centrosymmetric magnetic crystals. Dzyaloshinskii showed that, in this case, the DM interaction stabilises long-periodic spatially modulated structures of the vectors $\vec{S}_i$, structures with a fixed sense of rotation. In antiferromagnets, the DM-interaction favours arrangements of the magnetic moments, which results in a weak spontaneous magnetisation.

Within a continuum approximation for magnetic properties, the interactions responsible for these modulations are expressed by inhomogeneous invariants. In Ref.10, these contributions to the free magnetic energy, involving first derivatives of magnetisation with respect to spatial coordinates, are defined as the inhomogeneous Dzyaloshinskii-Moriya interactions. These interactions are linear with respect to first spatial derivatives of magnetisation $\vec{M}$ in an antisymmetric mathematical form, firstly studied in the theory of phase transitions by E. M. Lifshitz and known as Lifshitz invariants. Spiral structures arise in magnetic systems from the presence of the Lifshitz invariant in the free energy [11].

The structure of the Lifshitz invariant is, in the case of the inhomogeneous Dzyaloshinskii-Moriya interaction, a product of three vectors: a vector $\vec{D}$ representing an internal or external field or a fixed direction in the space, a vector $\vec{M}$ representing the local order parameter and the $\vec{\nabla}$ operator on the order parameter components. The product has the following form:

$$f_L = \vec{D} \cdot \left[ \vec{M} (\vec{\nabla} \cdot \vec{M}) - (\vec{M} \cdot \vec{\nabla}) \vec{M} \right] \qquad (1)$$

In the case of the liquid crystals, we shall see that vector $\vec{D}$ can be an external electric field or the direction perpendicular to the sample surface. It is better to remark that in Ref.4, we can find another choice for the DM interaction, as the pseudoscalar $f_{DM} = D[\vec{M} \cdot \vec{\nabla} \times \vec{M}]$. We will discuss this form in the Sect.5, concerning the chiral nematics.

We used the DM interactions in 1996, to study the field-induced phase transition of $BiFeO_3$ [12]. More recently, the coupling of spin waves with the optical phonons has been discussed in the framework of Lifshitz invariant, for the same material [13]. An antiferromagnetic vector $\vec{L}$ characterises the $BiFeO_3$ spin structure. The Landau-Ginzburg energy density [3] of the spin structure is the sum

$$f = f_L + f_{exch} + f_u + f_{ME} =$$
$$= -\alpha P_{z,S}(L_x \nabla_x L_z + L_y \nabla_y L_z) + A \sum_{i=x,y,z}(\nabla L_i)^2 - K_u L_z^2 - \beta \vec{E} \cdot \vec{H}_{DM} \quad (2)$$

The first term $f_L$ in Eq.2 is the magneto-electric coupling as a Lifshitz invariant, where $P_{z,S}$ is the z-component of the spontaneous polarization vector, and α is the inhomogeneous relativistic exchange constant (inhomogeneous magneto-electric constant). The Lifshitz invariant is the responsible for the spatially modulated spin structure in BiFeO$_3$, as shown in Ref.12. The second term $f_{exch}$ in (2) is the inhomogeneous exchange energy, where $A$ is a stiffness constant. In the third term, $K_u$ is the uniaxial anisotropy. $f_{ME}$ is the coupling of an external electric field $\vec{E}$ with a spatial uniform inner field $\vec{H}_{DM} = \vec{d} \times \vec{L}$, where $\vec{d} = (0,0,P_z)$, and β the homogeneous magneto-electric constant. This term is originated from a magneto-electric-like DM interaction.

The first term of the free energy can be rewritten, using the following vector:

$$\vec{A} = \vec{L}(\vec{\nabla} \cdot \vec{L}) - (\vec{L} \cdot \vec{\nabla})\vec{L} \quad (3)$$

in the form:

$$f_L = -\alpha \vec{P}_S \cdot \vec{A} \quad (4)$$

as a scalar product of two fields. In our paper [12], we investigated the influence of an electric field on the spatially modulated spin structure (SDW state). The electric field has a tendency to prefer a homogeneous state and to induce a phase transition to this state. In that paper, we used the analogy with nematic liquid crystals to study magnetic materials. Here, we want to enhance the analogy of liquid crystal interactions with the two form of the Dzyaloshinskii-Moriya DM interaction

**3. The flexoelectricity in liquid crystals**
Let us consider a nematic liquid crystal and assume as order parameter the director field $\vec{n}$, which is giving the local mean orientation of molecules. This is usually a unit vector. Vector $\vec{A}$ can then used in nematics too, rewritten in the following form:

$$\vec{A} = \vec{n}(\vec{\nabla} \cdot \vec{n}) - (\vec{n} \cdot \vec{\nabla})\vec{n} = \vec{n}\,\text{div}\,\vec{n} + \vec{n} \times \text{rot}\,\vec{n}\,. \quad (5)$$

Vector $\vec{A}$ in Eq.5 is well known in the physics of liquid crystals. $\vec{A}$ is encountered in the structure of flexoelectric contribution to bulk free energy as $f_{flexo} = -\vec{P} \cdot \vec{E}$. Flexoelectricity is a property of liquid crystals similar to the piezoelectric effect. In certain anisotropic materials, which contain molecular asymmetry or quadrupolar ordering with permanent molecular dipoles, an applied electric field may induce an orientational distortion. Conversely any distortion will induce a macroscopic polarization within the material. The polarization vector $\vec{P}$ in the flexoelectric term is then described with a distortion in the nematic director field:

$$\vec{P} = e_S \vec{n} (\vec{\nabla} \cdot \vec{n}) - e_B (\vec{n} \cdot \vec{\nabla}) \vec{n} = e_S \vec{n}\, div\, \vec{n} + e_B \vec{n} \times rot\, \vec{n}. \tag{6}$$

The two terms in the polarization vector are due to the splay and the bend contribution. The coupling of the polarization $\vec{P}$ with an external electric results in the appearance of a periodic distortion of an initial planar orientation of the nematic cell [14]. Meyer showed that the infinite liquid crystal must be disturbed, the perturbation is periodic along the director orientation and the period is inversely proportional to electric field strength [15]. This is not surprising because the polarization vector $\vec{P}$ has the same structure of vector $\vec{A}$ in Eq.5.

In the flexoelectricity, the polarization is induced by a deformation of the director field. Let us remember that in the piezoelectric materials, an applied uniform strain can induce an electric polarization or vice versa. Crystallographic considerations restrict this property to non centrosymmetric systems. A strain, which is not uniform, can potentially break the inversion symmetry and induce polarization in non piezoelectric materials. While the conventional piezoelectric property is different from zero only for certain select materials, the non local coupling of strain and polarization could be potentially found in all dielectrics [16]. As a result we find that the coupling with an external field gives the Lifshitz invariant as a DM non homogenous coupling for the electric field with the Lifshitz vector.

**4. Periodic distortions in nematics.**
Let us discuss more deeply the Meyer result [14,15] of a periodic distortion in the infinite medium. The free energy density is given by:

$$f = \frac{1}{2} K \left[ (div\, \vec{n})^2 + (\vec{n} \cdot rot\, \vec{n})^2 + (\vec{n} \times rot\, \vec{n})^2 \right] - e\, \vec{E} \cdot (\vec{n}\, div\, \vec{n} + \vec{n} \times rot\, \vec{n}) \tag{7}$$

in the uniform elastic approximation, with $K$ elastic constant, and with the dielectric anisotropy negligible. Moreover we assume $e_B \approx e_S \approx e$. Let us consider director $\vec{n}$ in a uniform configuration, as a vector parallel to $x$-axis and the electric field $\vec{E}$ parallel to $z$-axis as $\vec{E} = E\,\vec{k}$, where $\vec{k}$ is the unit vector of $z$-axis. Angles $\theta$ and $\phi$ are shown in the Figure 1.

The components of director $\vec{n}$ are $n_x = cos\,\theta$, $n_y = 0$, $n_z = sin\,\theta$, if $\phi = 0$. For the sake of simplicity, let us consider a deformation of $\vec{n}$ depending on $x$. In fact, we want to give just a rough approach to rigorous calculations. In the case of an infinite nematic medium without deformations of the director, the free energy density is zero. If we have a tilt angle variation of the form $\theta = -e\,xE/K$, we have a periodic deformation of director $\vec{n}$. The free energy density, including the flexoelectric term, is:

$$f_{distorted} = \frac{1}{2} K \left[ sin^2\theta \left(\frac{\partial \theta}{\partial x}\right)^2 + cos^2\theta \left(\frac{\partial \theta}{\partial x}\right)^2 \right] + eE \frac{\partial \theta}{\partial x} = -\frac{1}{2} e \frac{E^2}{K} < 0. \tag{8}$$

Then, a periodic distortion in a non-confined nematic is possible because it has a free energy density lower than that possessed by the uniform configuration. There is not a threshold for the electric field, since the existence of a threshold is a consequence of the medium confinement. Let us imagine a nematic material confined in a cell composed by two plane walls, parallel to [x,y] plane, at a distance $d$. The anchoring conditions must be included in the energy balance. We can assume a surface energy density of the Rapini-Papoular form $f = -W\cos^2\theta$, for a surface treatment favouring a molecular alignment parallel with the $x$-axis. If the director field $\vec{n}$ is uniform in the plana alignment, $f = -W$. Let us choose as in Ref.14, the behaviour of the tilt angle in the form $\theta = -\alpha xE/K$, with $\alpha$, a coefficient with dimensions $charge/length$. The free energy density is given as in Eq.8.

Integrating the free energy density on the cell volume $V = d \Lambda L$, where $d$ is the cell thickness, $L$ a fixed length in $y$-direction and $\Lambda$ the director distortion wavelength in $x$-direction, we obtain:

$$F_{distorted} = E^2 \frac{d\Lambda L}{K}\left[\frac{1}{2}\alpha^2 - e\alpha\right] - \frac{1}{2}W\Lambda L \tag{9}$$

The last term in (9) is the surface energy contribution. In the case of a uniform director field, we have a total energy as $F_{uniform} = -W\Lambda L$. The behaviour of the two free energies $F_{distorted}, F_{uniform}$ is given in Fig.2: we can see the existence of a threshold field $E^*$. If the electric field has a value $E < E^*$, the stable configuration of the director field is that with lower energy, in this case when the director field is uniform. When $E > E^*$, the stable configuration is the distorted one.

Comparing the two values of the total energy, that is:

$$F_{distorted} \approx F_{uniform} \tag{10}$$

we can approximately find the threshold electric field as:

$$(E^*)^2 \left(\frac{1}{2}\alpha^2 d - e\alpha d\right) = -\frac{W}{2}K \tag{11}$$

where $(\alpha^2 d/2 - e\alpha d) < 0$, to have a real electric field:

$$(E^*)^2 = \frac{W}{d}\frac{K}{|\alpha^2 - 2e\alpha|} \tag{12}$$

The threshold field has a value:

$$E^* = \left(\frac{WK}{\alpha d|\alpha - 2e|}\right)^{1/2} = \left(\frac{WK}{a^2 d}\right)^{1/2} \tag{13}$$

where $\alpha < 2e$. Estimating $a^2 = |\alpha(\alpha - 2e)| \approx e^2$ and assuming the parameter values $W = 10^{-6}\ J/m^2$, $K = 10^{-11}\ N$, $d = 10\ \mu m$, $a = 10^{-11}\ C/m$ we find a threshold voltage of $\approx 1\ Volt$.

## 5. The chiral nematic and the smectic phase.

Many researches have taken place in the field liquid crystals to find ferroelectric materials, from the earlier studies on the smectic phases till the more recent banana-like materials [16-19]. The smectic phases are organised in layers. There are three main smectic phases: A, C and C*. In the smectic C (SmC) phase the director $\vec{n}$ is tilted by a fixed angle, with respect to the layer normal $\vec{v}$. The chiral smectic (SmC*) phase shows in addition an intrinsic twist of the director from layer to layer. The symmetry breaking $C_{2h} \to C_2$ allows molecular electric dipoles to form a spontaneous electric polarization $\vec{P}$, which lies in the smectic planes. The macroscopic polarization vanishes in the SmC* phase, but an electric field parallel to the layers can distort the helicoidal structure, disfavouring SmC* and leading to a phase with a macroscopic polarization. In the Landau theory of smectic liquid crystals, the free energy is expanded in two order parameters: the projection $\vec{n}$ of the director onto the smectic layer plane and the layer polarization $\vec{P}$ [20]. The chiral term, responsible of the SmC* phase has in [20] the structure:

$$F_L = -D_2 \left( n_x \nabla_z n_y - n_y \nabla_z n_x \right) \tag{14}$$

This term has in fact a Lifshitz-like structure, if we consider the layer normal $\vec{v}$, parallel to the *z*-axis:

$$F_L = -D_2 (\vec{n} \times \vec{v}) \cdot \left[ \vec{v} (\vec{\nabla} \cdot \vec{n}) - (\vec{v} \cdot \vec{\nabla}) \vec{n} \right] = -D_2 (\vec{n} \times \vec{v}) \cdot \left[ \vec{v}\, div\,\vec{n} + \vec{v} \times rot\,\vec{n} \right]. \tag{15}$$

We can identify this expression as a pseudoscalar inhomogeneous Dzyaloshinskii-Moriya interaction, which does not involve an external field but a fixed direction in the space, that is the vector $\vec{v}$ normal to the smectic layer.

Chiral molecules can also form nematic phases called chiral nematic phases or cholesteric phases. The phase shows a nematic order but the director rotates throughout the sample. The axis of this screw is normal to the director. The distance over which the director rotates by $2\pi$ is the chiral pitch, generally of the order of the wavelength of visible light.

If the nematic phase is composed of chiral molecules, all of the same chirality, the material does not have symmetry planes and then the free energy has, according to Landau and Lifshitz, a pseudoscalar term:

$$F_{chiral} = b\,\vec{n} \cdot \vec{\nabla} \times \vec{n} \tag{16}$$

This is the pseudoscalar of the DM interaction as in Ref.4, and introduced in Sect.2. If we consider the vector $\vec{v}$ as the direction of pitch, then the director $\vec{n}$ lays in a plane perpendicular to it and then Eq.16 can be rewritten as:

$$F_{chiral} = b(\vec{n} \times \vec{v}) \cdot \left[ \vec{v}(\vec{\nabla} \cdot \vec{n}) - (\vec{v} \cdot \vec{\nabla}) \cdot \vec{n} \right], \tag{17}$$

with the same structure that we encountered in the smectic term originating the helix.

**6. The saddle-splay elasticity at surfaces.**
In nematic, a more general form of the distortion free-energy density, in the framework of the usual first-order continuum theory, is given as:

$$f = \frac{1}{2}K\left\{(div\,\vec{n})^2 + (\vec{n}\cdot rot\,\vec{n})^2 + (\vec{n}\times rot\,\vec{n})^2\right\} - (K+K_{24})\,div[\vec{n}\,div\,\vec{n} + \vec{n}\times rot\,\vec{n}] \quad (18)$$

where $K$ is the bulk elastic constant in the case of elastic isotropy. The last term is the contribution of the saddle-splay elasticity. This contribution is not usually inserted in the bulk free energy, because it is becomes a surface contribution when integration is performed on the cell thickness [21,22]. The saddle-splay contribution is then a Lifshitz invariant of the surface energy:

$$f_{surface} = -(K+K_{24})\vec{v}\cdot[\vec{n}\,div\,\vec{n} + \vec{n}\times rot\,\vec{n}] \quad (19)$$

if $\vec{v}$ is the unit vector of direction perpendicular to the surface containing the nematic material. This term has the same form of Lifshitz scalar product in Eq.4.
In addition to the anchoring energy, which is the anisotropic part of surface tension, there is an elastic contribution, which has been originally indicated as a part of the bulk elastic energy in the form of a divergence [23-25]. This contribution can be viewed as the elastic part of surface energy depending on the tangential gradient of director. The $K_{24}$ term may induce spontaneous twist deformations in hybrid nematic films with azimuthally degenerate anchoring conditions. Such deformations are manifested in the formation of periodic stripe domains observed in sufficiently thin hybrid NLC cells [21,26].
The saddle-splay contribution is necessary, if we have to evaluate the elastic contribution of thin films or membranes. In 1973, Helfrich studied the energetic cost of a generic sheet in a three-dimensional space: we can determine, in each point of the sheet, the radii of curvature $r_1$ and $r_2$, and local curvatures $c_1 = 1/r_1$ and $c_2 = 1/r_2$. Curvatures can be positive or negative. Saddle-shaped surfaces have curvature that is positive along one principle axis and negative along the other. The energetic cost per unit area associated with bending a membrane, as noted by Helfrich [27], is given by the sum of two terms, one dependent on total curvature, $c_1 + c_2$, and the other on product $c_1 c_2$:

$$F_S = \frac{1}{2}k_C(c_1 + c_2 - 2c_o)^2 + k_G c_1 c_2 \quad (20)$$

In this expression, $k_C$ is the bending (or curvature) modulus and $k_G$ is the saddle-splay (or Gaussian curvature) modulus. These two modules are set by interactions among membrane molecules. The spontaneous curvature is denoted by $c_o$. As reported in [28], biological membranes are sheets that can be modelled with a continuum elastic approach. These membranes are two-dimensional fluids within which proteins diffuse and interact. Membranes can bend and curve, with deformations controlled by proteins and lipids; the converse is also true, that it the structure created by membrane curvature can guide the spatial organisation of membrane molecules. Then the membrane can display spatial patterning at length-scales far greater than the scale of individual molecules [28].

## 7. The hybrid cell and the flexoelectricity

We start here the discussion of some case studies. The first is on the role of flexoelectricity in the hybrid nematic cells. Let us remember that the hybrid cell is a nematic cell, where a sample is confined between two parallel walls with different anchoring conditions. One surface is treated to favours a planar alignment; the opposite one is favouring a homeotropic alignment. The cell is then named HAN, that is Hybrid Aligned Nematic cell. The hybrid cell we discuss has the *y*-axis perpendicular to cell walls (see the upper part of Fig.3). An electric field can be applied parallel to *y*-axis: we have then $\vec{E} = E\vec{j}$ where $\vec{j}$ is the unit vector of *y*-axis. $\vec{j}$ is the homeotropic direction too. The unit vector $\vec{i}$, parallel to the cell walls, gives the easy planar direction. The bulk free energy density is given, in the elastic isotropic approximation, by:

$$f_{bulk} = \frac{1}{2} k \left[ (div\,\vec{n})^2 + (rot\,\vec{n})^2 \right] - \frac{\varepsilon_o \Delta\varepsilon}{2} (\vec{E} \cdot \vec{n})^2 \qquad (21)$$

where the last term is due to the dielectric anisotropy $\Delta\varepsilon$ of the nematic. The surface energy in the Rapini-Papoular can be used:

$$f_{Surf} = \left[ W_P (\vec{n} \cdot \vec{i})^2 - W_H (\vec{n} \cdot \vec{j})^2 \right], \qquad (22)$$

at the two surfaces, for $y = d$ and for $y = 0$. $W_P, W_H$ are energy densities of the surface anchoring. If we have a planar cell with surface *S*, thickness *d*, and a uniform director configuration $\vec{n} = n\vec{i}$, the total free energy is $F_{Planar} = -2S W_P$. If the director configuration is uniform but homeotropic, then $\vec{n} = n\vec{j}$ and the total free energy is the sum of the energy due to the presence of electric field and surfaces: $F_{Hom} = -\varepsilon_o \Delta\varepsilon E^2 / 2 - 2SW_H$.

When $W_P, W_H > 0$, we have a homeotropic cell; if $W_P, W_H < 0$ the cell is planar. Graphically comparing (lower part of Fig.3) the energies of the homeotropic and planar cells, we see the possibility of an electric threshold field $E^*$: under this value of the electric field, it is favoured the planar configuration, over the threshold value, it is the homeotropic configuration that has a lower energy.

In a hybrid cell, the director changes from a planar configuration at one of the cell wall, to a homeotropic configuration at the other cell wall. The tilt angle is then depending on *y*, as a function $\theta = \theta(y)$. The director field is: $\vec{n} = cos\,\theta\,\vec{i} + sin\,\theta\,\vec{j}$. If the anchoring is strong, the tilt angle is $\theta = \pi/2$ at $y = 0$ - homeotropic wall, and $\theta = 0$ at $y = d$ - planar wall. In the one elastic constant approximation, we have as the bulk free energy density:

$$f_{Bulk} = \frac{K}{2} \left( \frac{\partial \theta}{\partial y} \right)^2 - \frac{\varepsilon_o \Delta\varepsilon}{2} E^2 \,sen^2\theta \qquad (23)$$

and the surface energy density $f_{Surf} = -W_P - W_H$. As hybrid configuration, let us simply choose a linear function of the tilt angle with *y*. Then:

$$\theta = -\frac{\pi}{2}\frac{y}{d} + \frac{\pi}{2} \tag{24}$$

with $\theta_o = \pi/2$ and $\theta_d = 0$. Then $\partial\theta/\partial y = -\pi/2d$ and the total bulk energy is:

$$F_{Bulk} = S\int_0^d \frac{K}{2}\left(\frac{\partial\theta}{\partial y}\right)^2 dy - \frac{\varepsilon_o \Delta\varepsilon}{2} E^2 S \int_0^d sen^2\theta\, dy = \frac{K\pi^2 S}{8d} - \frac{\varepsilon_o \Delta\varepsilon}{4} E^2 S d \tag{25}$$

Including the surface energy, the total energy is:

$$F_{Tot} = \frac{K\pi^2 S}{8d} - \frac{\varepsilon_o \Delta\varepsilon}{4} E^2 S d - S(W_P - W_H) \tag{26}$$

Let us compare this expression with the energy of the cell in homeotropic and planar configurations, choosing an anchoring energy favouring planar and hybrid configurations under threshold fields:

$$F_{Planar} = -2SW_P \quad ; \quad F_{Homeotropic} = -\frac{\varepsilon_o \Delta\varepsilon}{2} E^2 S d - 2SW_H$$

$$F_{Hybrid} = \frac{K\pi^2 S}{8d} - \frac{\varepsilon_o \Delta\varepsilon}{4} E^2 S d - S(W_P - W_H) \tag{27}$$

What is shown in Fig.4(a) is possible, because we can adjust the anchoring parameters. We observe then two threshold fields: when the field is lower than $E'$, the nematic is planar, if the field is comprised between $E'$ and $E''$, the cell is hybrid. Over $E''$, the cell is homeotropic.
As previously discussed, the electric field can be coupled with a polarization arising from an elastic deformation in the flexoelectric effect. In planar and homeotropic configurations, because there are not deformations of director, the flexoelectric effect is absent, but in the hybrid cell the deformation gives a flexoelectric polarization $\vec{P} = (e_S\, \vec{n}\, div\,\vec{n} + e_B\, \vec{n}\times rot\,\vec{n})$, different from zero.
Let us add the term $f_{flexo} = -\vec{P}\cdot\vec{E}$ to the free energy density, which is

$$f_{flexo} = -\vec{P}\cdot\vec{E} = -(e_S - e_B) E\, sen\theta\, cos\theta\, \frac{\partial\theta}{\partial y}. \tag{29}$$

If $\theta$ is given by Eq.(24), after integrating on the cell volume, we have the contribution of flexoelectricity to the total free energy as:

$$F_{Flexo} = -(e_S - e_B) E\, S \tag{30}$$

In principle, coefficient $(e_S - e_B)$ could be positive or negative, depending on the value of splay and bend parameters. The threshold values $E', E''$ are changed from the contribution of the flexoelectricity. They could be lowered or raised by the induced polarization (see Fig.4(b)). The thresholds change according to the shape of the molecules. Comparing the thresholds we can estimate the values of the coefficients. The two electric field contributions in the HAN cell are:

$$F_1 = -\frac{\varepsilon_o \Delta\varepsilon}{4} E^2 S d \quad ; \quad F_2 = -(e_S - e_B) E S \tag{31}$$

If they were of the same order of magnitude, we could obtain:

$$(e_S - e_B) \approx \frac{\varepsilon_o \Delta\varepsilon}{4} E d \tag{32}$$

If the cell has a thickness of $10\,\mu m$ and the field has the value of $10\,V/\mu m$, and an electric anisotropy as $\Delta\varepsilon = 0.1$ then:

$$(e_S - e_B) \approx 25 \frac{pC}{m} \tag{33}$$

in agreement with Ref.29 and with other experimental values [30-34]. Recently a giant flexoelectricity has been found with bent-core nematics: a peak of $35\,nC/m$ was measured in these materials then more than 3 orders of magnitude larger than in calamitics [35]. In the next section we will study the alignment transitions in the nematic cells; such a problem was studied also in Ref.36.

**8. The phase diagrams of the hybrid cell**
Let us consider the hybrid cell as in the previous section. We use the same notation here but we solve the Euler-Lagrange equation with the proper boundary conditions, by means of an iterative procedure previously used in Ref.37, to investigate the ion densities in corona plasma. The Euler-Lagrange is:

$$\frac{\partial f_{bulk}}{\partial \theta} = \frac{\partial}{\partial y} \frac{\partial f_{bulk}}{\partial (\partial \theta/\partial y)} \tag{34}$$

and

$$f_{bulk} = \frac{K}{2}\left(\frac{\partial \theta}{\partial y}\right)^2 - \frac{\varepsilon_o \Delta\varepsilon}{2} (E \sin\theta)^2 \tag{35}$$

that is:

$$\frac{\partial^2 \theta}{\partial y^2} + \xi^2 \sin\theta \cos\theta = 0 \quad ; \quad \xi^2 = \varepsilon_o \Delta\varepsilon E^2 / K \tag{36}$$

and the surface energy density:

$$f_{surf} = -W_H \sin^2\theta_o + W_P \sin^2\theta_d \tag{37}$$

where , and $W_P > 0$. The boundary conditions are given by the following equations:

$$-K\left(\frac{\partial \theta}{\partial y}\right)_{y=0} = |W_H|(sin\,2\theta)_{y=0} \quad ; \quad -K\left(\frac{\partial \theta}{\partial y}\right)_{y=d} = |W_P|(sin\,2\theta)_{y=d} \tag{38}$$

We choose a solution of the form:

$$\theta = \theta^o(y) + \theta'(y) \tag{39}$$

and then Eq.36 can be written as two equations:

$$\frac{\partial^2 \theta^o}{\partial y^2} = 0 \quad ; \quad \frac{\partial^2 \theta'}{\partial y^2} = -\xi^2 sin\,2\theta^o \tag{40}$$

The second equation in (40) is solved in the following iteration:

$$\frac{\partial^2 \theta^{j+1}}{\partial y^2} = -\xi^2 sin\,2\tilde{\theta}^j \tag{41}$$

where $\tilde{\theta}^j = \theta^o + \theta^j$ and $\tilde{\theta}^o = \theta^o$. Three steps of the iteration are enough to have the solution. Then $\theta = \theta^o(y) + \theta^{j+1}(y)$ where:

$$\theta^o(y) = \alpha_o + \alpha_1 y \quad ; \quad \theta^{j+1}(y) = -\xi^2 \int_0^y \int_0^{y'} sin\,2\tilde{\theta}^j(y')dy'\,dy'' \tag{42}$$

The boundary conditions are:

$$d\left(\frac{\partial \theta}{\partial y}\right)_{y=0} + b_o(sin\,2\theta)_{y=0} = 0 \quad ; \quad d\left(\frac{\partial \theta}{\partial y}\right)_{y=d} + b_1(sin\,2\theta)_{y=d} \tag{43}$$

in which we used the dimensionless parameters $|W_H d/K| = b_o$ ; $|W_P d/K| = b_1$. From the first equation in the boundary conditions (43):

$$\alpha_1 d + b_o(sin\,2\alpha_o) = 0 \tag{44}$$

Once we choose the value of $\alpha_o$, from Eq.(44), we have the value of $\alpha_1$, and then, after iteration, the solution $\theta = \theta^o + \theta'$. To determine the value of $\alpha_o$ we could use the other boundary condition, the second in (43); but, in this case, we are facing a strongly oscillating function. It is better to determine the value of parameter $\alpha_o$, minimizing the total free energy. Adding the flexoelectricity, the term to include in the free energy density is:

$$f_{flexo} = -(e_S - e_B)E \sin\theta \cos\theta \frac{\partial \theta}{\partial y} \tag{45}$$

and, after integration on the cell thickness, we have a further contribution to the surface energy density of the form:

$$F_{flexo} = -(e_S - e_B)E[\sin\theta_o - \sin\theta_d] \tag{46}$$

This term can be easily inserted in the numerical calculation, to minimize the total free energy. Let us introduce the following dimensionless variables and parameter:

$$\hat{y} = \frac{y}{d} \; ; \; \hat{\xi} = \xi d \; ; \; \Pi = (e_S - e_B)\sqrt{\frac{2}{K\varepsilon_o \Delta\varepsilon}} \tag{47}$$

to illustrate the results of calculations. In Fig.5 we can see the phase diagrams of the HAN cell, for a fixed choice of the surface parameter $b_o = 1$. We can change the value of parameter $b_1$ and find the value of the threshold field (the electric field is dimensionless represented by $\hat{\xi}$). There are three regions in the diagrams where planar, homeotropic and hybrid alignments are allowed according to the values of the electric field. The phase diagram is depending on the values of flexoelectric parameter $\Pi$ (see diagrams (a),(b) and (c) in Fig.5). The last diagram (d) shows the behaviour of a cell when we change the flexoelectric parameter $\Pi$. Note that the hybrid configuration disappears when flexoelectric parameter is higher than value 1.3.

In the Fig.6, we see the behaviour of $\cos\theta$ as a function of the dimensionless variable $\hat{y} = y/d$ in the case of positive and negative flexoelectric coefficients, for different values of the electric field. Note that, as the field increases, the role of surface is suppressed and the angle at the planar surface increases. As the electric field is higher than the threshold value, the cell becomes homeotropic and $\cos\theta = 0$.

To conclude this section on HAN cell, let us remember that we have another Lifshitz invariant, that giving the saddle-splay contribution to the surface free energy density, in the form: $f_{Saddle-splay} = -(K + K_{24})\vec{v} \cdot [\vec{n}\, div\,\vec{n} + \vec{n} \times rot\,\vec{n}]$. In the hybrid cell alignment, where only the tilt angle is displayed by the elastic distortion, this contribution is zero. We will see in the last section of the paper, how this term is production periodic distortion and how the PHAN - the Periodic HAN - texture appears.

The fact that the saddle-splay contribution is zero, in the HAN configuration, is in agreement with the conclusion that in the same configuration the flexoelectric contribution $f_{flexo} = -\vec{E} \cdot [e_S \vec{n}\, div\,\vec{n} + e_B \vec{n} \times rot\,\vec{n}]$ is zero too, when $e_S = e_B$. As we saw in Sect.4, it is the periodic distortion to origin a contribution different from zero, if $e_S = e_B$.

**9.Nematics in the cylindrical geometry.**
Let us consider a cylinder with radius $R$. In this cylindrical cell we imagine to insert a nematic. We use the frame as in Fig.7 and solve the Euler-Lagrange equation in cylindrical coordinates. Let us consider $\theta = \theta(r)$, only depending on the radial distance, and moreover, $\phi = 0$. The Euler-Lagrange is:

$$\frac{\partial f_{bulk}}{\partial \theta} = \frac{1}{r}\frac{\partial}{\partial r}\left(r\frac{\partial f_{bulk}}{\partial(\partial\theta/\partial r)}\right) \tag{48}$$

The bulk density energy is given by:

$$f_{bulk} = \frac{K}{2}\left[\left(\frac{\partial\theta}{\partial r}\right)^2 + \frac{\sin^2\theta}{r^2} + 2\frac{\sin\theta\cos\theta}{r}\left(\frac{\partial\theta}{\partial r}\right)\right] - \frac{\varepsilon_o\Delta\varepsilon}{2}(E\cos\theta)^2 \tag{49}$$

and then the Euler-Lagrange equation turns out to be:

$$\frac{\sin\theta\cos\theta}{r^2} + \xi^2\sin\theta\cos\theta = \frac{1}{r}\frac{\partial}{\partial r}\left(r\frac{\partial\theta}{\partial r}\right) \quad;\quad \xi^2 = \varepsilon_o\Delta\varepsilon E^2/K \tag{50}$$

The surface energy density is:

$$f_{surf} = -W_H\sin^2\theta_R + W_P\sin^2\theta_R \tag{51}$$

where $W_H > 0$; $W_P < 0$. For an anchoring, which favours an homeotropic alignment of the nematic perpendicular at the wall of the cylinder, we use:

$$f_{surf} = -W_H\sin^2\theta_R \tag{52}$$

If we want to avoid the presence of a defect at the axis of cylinder ($z$ – axis), the director must escape in the $z$ – direction. The solution, if the applied electric field is zero, is given by an inverse tangent:

$$\theta^o(r) = 2\tan^{-1}\left(\beta\frac{r}{R}\right) \tag{53}$$

where $\beta = 1$, for a strong anchoring at the cylinder wall. To solve the equation in the case of electric field different from zero, we choose a solution as:

$$\theta(r) = \theta^o(r) + \theta'(r) \tag{54}$$

Then we have two equations to solve:

$$\begin{aligned}\frac{\sin\theta^o\cos\theta^o}{r^2} &= \frac{1}{r}\frac{\partial}{\partial r}\left(r\frac{\partial\theta^o}{\partial r}\right)\\ \xi^2\sin\theta^o\cos\theta^o &= \frac{1}{r}\frac{\partial}{\partial r}\left(r\frac{\partial\theta'}{\partial r}\right)\end{aligned} \tag{55}$$

The second equation can be solved with iterations and three steps are enough. In the following way, we have:

$$\xi^2 \sin\widetilde{\theta}^j \cos\widetilde{\theta}^j = \frac{1}{r}\frac{\partial}{\partial r}\left(r\frac{\partial \theta^{j+1}}{\partial r}\right) \quad ; \quad \widetilde{\theta}^j = \theta^o + \theta^j \tag{56}$$

Actually, we arrive at solutions:

$$\theta^o(r) = 2\tan^{-1}\left(\beta\frac{r}{R}\right) \quad ; \quad \theta^{j+1}(r) = \xi^2 \int_0^r \frac{dr''}{r''}\int_0^{r''} \sin\widetilde{\theta}^j(r')\cos\widetilde{\theta}^j(r') r' dr' \tag{57}$$

and then at final solution $\theta(r) = \theta^o(r) + \theta^{j+1}(r)$. To determine the value of parameter β we choose the solution minimizing a reduced total free energy:

$$\frac{F_{bulk}}{2\pi RL} = \int_0^d f_{bulk}\, r\, dr + f_{surf} \tag{58}$$

where $L$ is an arbitrary length of the cylindrical cell. Let us then consider the contribution of flexoelectricity to Euler-Lagrange equations:

$$\frac{\partial f_{flexo}}{\partial \theta} - \frac{1}{r}\frac{\partial}{\partial r}\left(r\frac{\partial f_{flexo}}{\partial(\partial\theta/\partial r)}\right) = (e_S - e_B)E\frac{\sin^2\theta}{r} \tag{59}$$

We use again $(e_S - e_B)E = \Pi\xi$ and $b = WR/K$ as parameters. The equation to solve is:

$$\xi^2 \sin\widetilde{\theta}^j \cos\widetilde{\theta}^j + \Pi\xi\frac{\sin^2\widetilde{\theta}^j}{r} = \frac{1}{r}\frac{\partial}{\partial r}\left(r\frac{\partial \theta^{j+1}}{\partial r}\right) \tag{60}$$

instead of Eq. (56). The Fig.8 and Fig. 9 report the results of calculations for different values of anchoring and flexoelectric coefficients. In the Fig.8 we can see the angle θ as a function of the reduced radial distance $r/R$, for two values of the flexoelectric coefficient, Π = 0 and 1. The figure shows the behaviour in the case of different values of anchoring parameter $b$ and of dimensionless electric field parameter $\hat{\xi}$. As the electric field is higher that a threshold value, angle θ goes to zero and the director field is parallel to cylinder axis in all the cell. The following Fig.9 shows $\cos\theta$ as a function of reduced radial distance $r/R$, for different values of Π and $\hat{\xi}$. In this case, the value of the anchoring strength is fixed. Note that a negative value of the flexoelectric parameter is strongly favouring the alignment of the director parallel to cylinder axis, and then we find a low value of the threshold electric field. If flexoelectric parameter Π is positive and large, the distorted configuration is favoured, and the threshold field required for suppressing this configuration is increased. Moreover, if the flexoelectric parameter is large, as in the lower image in Fig.9, angle θ starts to oscillate as the field increases. We must have a huge electric field to suppress the oscillating distortion and have $\cos\theta = 0$, with all the nematic aligned parallel to the field, in a uniform configuration. In Fig.10, the phase diagrams, when anchoring parameter $b$ is fixed and equal to 6. We see three regions, denoted by: U for uniform alignment of director parallel to *z*-axis, D if the director has a deformed configuration, and O when the director is oscillating and cosine becomes negative too. Angle θ turns more than $\pi/2$ on the distance *R*. As told before, giant flexoelectric coefficients are possible and then the oscillation could be experimentally tested in cylindrical cells.

A last note on the flexoelectric term. The flexoelectric vector is a sum of two contributions:

$$\vec{P} = (e_S \, \vec{n} \, div \, \vec{n} + e_B \, \vec{n} \times rot \, \vec{n}) =$$
$$= e_s D(sin\theta \vec{u}_r + cos\theta \vec{u}_z) + e_B R(-cos\theta \vec{u}_r + sin\theta \vec{u}_z) = e_S D \vec{n} + e_B R \vec{t} \quad (61)$$

where $D = div \, \vec{n}, R = (rot \, \vec{n})_\phi$. These two component which are perpendicular each other: when they are coupled with the electric field, parallel to the cylinder axis, we have then the two contributions in bulk energy with an opposite sign.

To conclude this section, let us discuss the saddle-splay contribution to the free energy, that is:

$$f_{Saddle-splay} = -(K + K_{24}) \, div[\vec{n} \, div \, \vec{n} + \vec{n} \times rot \, \vec{n}] \quad (62)$$

In the previous assumptions, $\theta = \theta(r), \phi = 0$, $f_{Saddle-splay}$ is simply renormalizing the value of the surface energy and then we do not further discuss it.

**10. The saddle-splay contribution and the PHAN cell**

Sometimes, it is possible to note a periodicity in the HAN cells observed by the polarised light microscope. Because of this periodic configuration, the cell is in the PHAN configuration, that is a nematic cell with a period hybrid alignment. Two angles describe the PHAN configuration: θ and φ. The last angle is formed by the projection of the director in the plane of the cell with the *x*-axis.

The free energy density is that of Nehring and Saupe, and given by Eq.(19). The frame of reference is $[xyz]$, with $[xy]$ the cell plane and $z$ – axis perpendicular to the cell plane. The homeotropic wall is at $z_0 = 0$, where *z* is the axis perpendicular to the cell plane. The planar wall is at $z_1 = d$, where *d* is the thickness of the cell. The easy-axis of the planar alignment is chosen coincident with the *x*-axis. The director $\vec{n}$ is described as:

$$\vec{n} = \vec{i} \, cos\phi \, cos\theta + \vec{j} \, sin\phi \, cos\theta + \vec{k} \, sen\theta \quad (63)$$

The Euler-Lagrange equations are non-linear. They were solved in Ref.21, with a numerical approach to determine the threshold thickness of the cell between the planar and the PHAN. Here, we want to grasp the role of the saddle-splay contribution, with just simple calculations. Let us then consider the tilt angle θ depending on z, and the φ angle depending on x, in the following way:

$$\theta(z) = \frac{\pi z}{2d} \quad ; \quad \phi(x) = \frac{2\pi x}{\Lambda} \quad (64)$$

The tilt is zero if $z = 0$, and it is $\pi/2$ at $z = d$. With $\Lambda$ we denote the wavelength along $x$–axis. The free energy density is:

$$f_{Elastic} = \frac{K}{2}\left[\left(\frac{\pi}{2d}\right)^2 + \left(\frac{2\pi}{\Lambda}\right)^2 \sin^2\phi - \frac{2\pi^2}{\Lambda d}\cos^2\theta \sin\phi\right] \tag{65}$$

Let us integrate on the volume $V = d\Lambda D$, where $D$ is a fixed distance on $y$–axis, we have:

$$F_{Elastic} = \frac{K}{2}\Lambda D\, d\left[\left(\frac{\pi}{2d}\right)^2 + \frac{1}{2}\left(\frac{2\pi}{\Lambda}\right)^2\right] \tag{66}$$

Neglecting the anchoring with respect to $\phi$, and assuming just tilt anchoring, with a surface energy density of the form:

$$f_{Surf} = -W\, n_z^2 \tag{67}$$

where $W = W_P$ for planar anchoring with $\theta = 0$, and $W = W_H$ at the homeotropic anchoring $\theta = \pi/2$. After integrating on surfaces of the cell:

$$F_{Surf} = -(W_P + W_H)\Lambda D \tag{68}$$

and then the total free energy is:

$$F = \frac{K}{2}\Lambda D d\left[\left(\frac{\pi}{2d}\right)^2 + \frac{1}{2}\left(\frac{2\pi}{\Lambda}\right)^2\right] - (W_P + W_H) D \Lambda \tag{69}$$

Let us evaluate the saddle-splay contribution to free energy density, using Eq.4 of Ref.21, that here reduces to:

$$f_{Saddle-splay} = 2(1+\kappa_4)K\left[-\theta_H\left(\frac{\partial\varphi}{\partial x}\right)_H\right] = 2(1+\kappa_4)K\left[-\frac{\pi}{2}\frac{2\pi}{\Lambda}\right] \tag{70}$$

where $\kappa_4 = K_{24}/K$; after integration on a surface $S = \Lambda D$, we have:

$$F_{Saddle-splay} = -2(1+\kappa_4)K\pi^2 D \tag{71}$$

The total energy is then:

$$F_{PHAN} = \frac{K\pi^2 D\Lambda}{8d} + \frac{K\pi^2 d D}{\Lambda} - (W_P + W_H)\Lambda D - 2(1+\kappa_4)K\pi^2 D \tag{72}$$

Comparing with the free energy of HAN configuration:

$$F_{PHAN} \approx F_{HAN} \qquad (73)$$

and after simple calculations we find:

$$d^2 - 2(1+\kappa_4)\Lambda d - \frac{1}{8}\Lambda^2 = 0 \qquad (74)$$

Neglecting the last term, we find a threshold value for the cell thickness:

$$d_c \cong 2\Lambda(1+\kappa_4) \qquad (75)$$

If $d > d_c$, then we find a HAN configuration, but if $d < d_c$ the modulated PHAN texture is displayed in the cell. In Ref.21, we can see the experimental observation of thickness threshold in a nematic sample. This is just a rough discussion on the role of saddle-splay contribution in producing periodic instabilitis, but enough to understand the origin of a threshold thickness in the sample.

Let us remember that $f_{Saddle-splay} = -(K+K_{24})\vec{v}\cdot[\vec{n}\,div\,\vec{n} + \vec{n}\times rot\,\vec{n}]$ is a Lifshitz invariant, with the same structure of flexoelectric contribution $f_{flexo} = -\vec{E}\cdot[e_S\vec{n}\,div\,\vec{n} + e_B\vec{n}\times rot\,\vec{n}]$ when $e_S = e_B$. We could imagine a surface contribution of the form $f_{Saddle-splay} = -\vec{v}\cdot[c_S\vec{n}\,div\,\vec{n} + c_B\vec{n}\times rot\,\vec{n}]$, where coefficients are different. This could increase the variety of observable configurations.

**Conclusion.**
This paper is divided in two parts. In the first we discussed the analogies among Lifshitz invariants in magnetic materials and liquid crystals. We saw that the structure of these invariants is the same, and that they are producing periodic instabilities in both cases. In the Lifshitz invariant, the interaction is between an external action and the order parameter, in a form that contains the gradient of order parameter. The external action can be an electric field applied to the bulk, and in this case the relevant effect is the flexoelectricity, or the confinement due to free surfaces or cell walls. The Lifshitz invariant related to surfaces gives the saddle-splay contribution to surface energy.
In the second part of the paper we deeply discussed the role of flexoelectricity in case of confined nematics. We performed detailed calculations in the case of planar and cylindrical geometries. Phase diagrams are also shown, to see the alignment phase transitions due to electric field and the role of flexoelectric parameter.

# References

[1] L. D. Landau and E. M. Lifshitz, *Statistical Physics*, Pergamon Press, Oxford, 1980.
[2] A. Ya. Braginsky, Phys. Rev. B **66**, 054202, 2002.
[3] I.E. Dzyaloshinskii, Sov. Phys. JETP **19**, 960, 1964.
[4] A.N. Bogdanov, U.K. Rössler and C. Pfleiderer, Physica B **259-361**, 1162, 2005.
[5] A..M. Kadomtseva, Yu.F. Popov, G.P. Vorob'ev, V.Yu. Ivanov, A.A. Mukhin and A.M. Balbashov, JETP Letters **81**, 590, 2005.
[6] I. Dozov, Europhys. Lett. **56**, 247, 2001.
[7] Sarkissian, J. B. Park, B. Ya. Zeldovich, N. V. Tabirian, Quantum Electronics and Laser Science Conference, QELS '05, Vol.3, pp.1597 – 1599, 2005.
[8] T. Moriya, Phys. Rev. **120**, 91, 1960.
[9] D. Coffey. T.M. Rice and F.C. Zhang, Phys. Rev. B. **44**, 10112, 1991.
[10] A.N. Bogdanov, U.K. Rössler, M. Wolf, and K.-H. Müller, arXiv:cond-mat/0206291, 2002.
[11] E.M. Lifschitz, JETP **11**, 253, 1941.
[12] A.Sparavigna, A.Strigazzi, A.K.Zvezdin, Phys. Rev. B **50**, 2953, 1994.
[13] R de Sousa and J. E. Moore, Optical coupling to spin waves in the cycloidal multiferroic BiFeO$_3$, arXiv:0706.1260v2 [cond-mat.str-el], 2007.
[14] Y.P. Bobylev, V.G. Chigrinov ans S.A. Pikin, J. de Physique **40**, 331, 1979.
[15] R.B. Meyer, Phys. Rev. Lett. **22**, 918, 1969.
[16] R. Maranganti, N. D. Sharma, and P. Sharma, Phys. Rev. B **74**, 014110, 2006.
[17] S.T. Lagerwall, Ferroelectrics **301**, 15, 2004.
[18] N.A Clark, S.T. Lagerwall, Appl. Phys. Lett. **36**, 899, 1980.
[19] W. Weissflog, Ch. Lischka, S. Diele, G. Pelzl, I. Wirth, S. Grande, H. Kresse, H. Schmalfuss, H. Hartung, A. Stettler, Mol. Cryst. Liq. Cryst. **333**, 203, 1999.
[20] L. Benguigui and A.E.Jacobs, Phys. Rev. E **49**, 4221, 1994.
[21] A.Sparavigna, O.D. Lavrentovich and A.Strigazzi, Phys. Rev. E 49, 1344, 1994.
[22] G.Barbero, A.Sparavigna and A.Strigazzi, Nuovo Cimento D **12**, 1259, 1990.
[23] F.C. Frank, Discuss. Faraday Soc. **25**, 19, 1958.
[24] T.C. Lubensky, Phys. Rev. A **2**, 2497, 1970.
[25] J. Nehring and A. Saupe, J. Chem. Phys. **54**, 337, 1971.
[26] V. M. Pergamenshchik, Phys. Rev. E **47**, 1881, 1993.
[27] W. Helfrich, Z. Naturforsch. **28c**, 693, 1973.
[28] R.Parthasarathy and J.T. Groves, Soft Matter **3**, 24, 2007.
[29] D. Lai Gwai Cheung, Structures and Properties of Liquid Crystals and Related Molecules from Computer Simulation, Ph.D Thesis, University of Durham, 2002.
[30] D. Schmidt, M. Schadt, and W. Helfrich, Z. Naturforsch. **27a**, 277, 1972.
[31] G. Barbero, P. T. Valabrega, R. Bartolino, and B. Valenti, Liq. Cryst. **1**, 483, 1986.
[32] I. Dozov, Ph. Martinot-Lagarde, and G. Durand, J. Phys. (Paris), Lett. **43**, L-365, 1982.
[33] S. Warrier and N.V. Madhusudana, J. Phys. II (France) **7**, 1789, 1997
[34] L. M. Blinov *et al.*, Phys. Rev. E **64**, 031707, 2001.
[35] J. Harden, B. Mbanga, N. Èber, K. Fodor-Csorba, S. Sprunt, J. T. Gleeson, A. Jákli, Phys. Rev. Letters **97,** 157802, 2006.
[36] R. Barberi, G. Barbero, Z. Gabbasova and A. Zvezdin, Phys. II France **3**, 147, 1993.
[37] A. Sparavigna and R.A. Wolf, Czechoslovak Journal of Physics **56**, B1062, 2006.

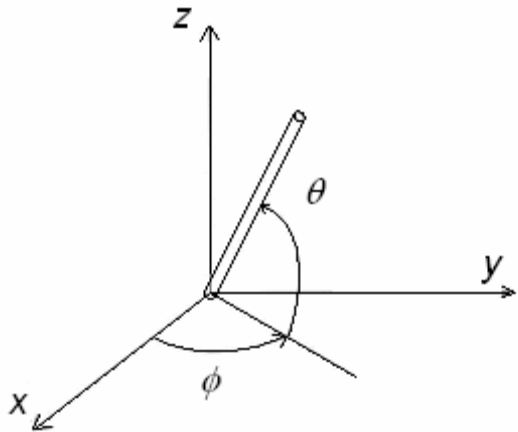

*Fig.1 The frame of reference and the angles used to describe the director, represented by the rod-like molecule.*

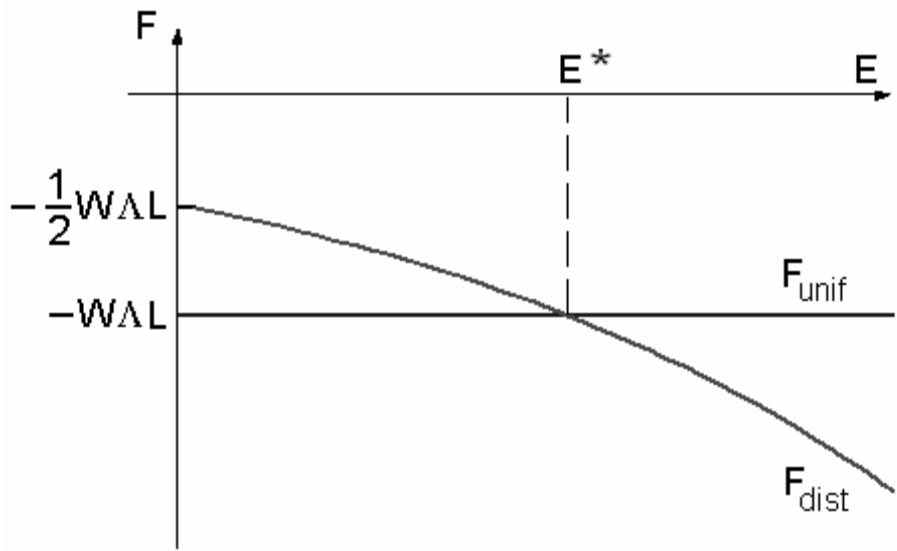

*Fig.2 Comparison of the free energy behaviours in the case of the uniform configuration and for the distorted one.*

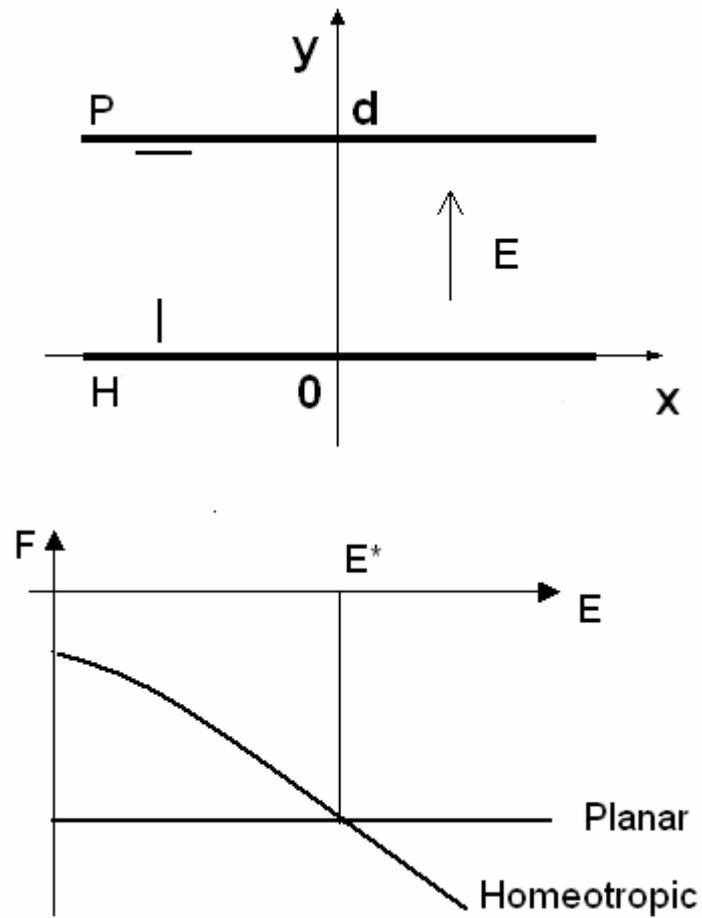

*Fig.3 Frame of reference for the hybrid cell in the upper part of the figure. In the lower part, the free energies as a function of the electric field in the planar and in the homeotropic configuration. Note the presence of a threshold.*

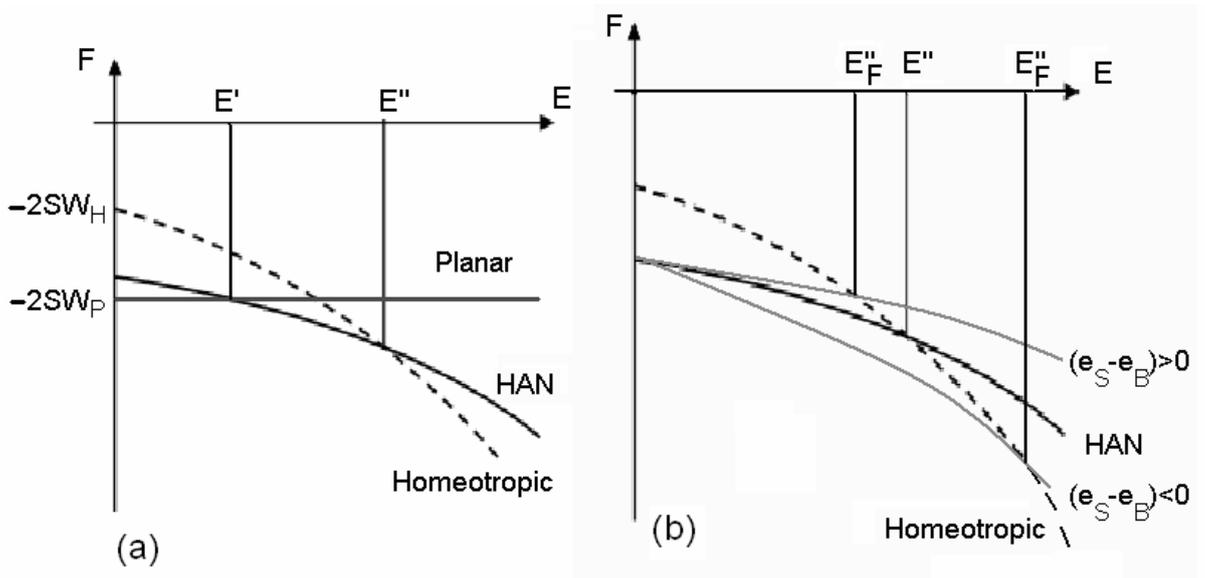

*Fig.4 The behaviour of the free energies of planar, hybrid (HAN) and homeotropic configuration, as functions of the electric field (a). Note the existence of two thresholds for the transition between the planar and the HAN configuration and between the HAN and the homeotropic configuration. In (b), the two curves in grey show how the energy of the HAN configuration changes for the presence of flexoelectricity. According the sign of the flexoelectric parameter, the threshold field is raised or lowered.*

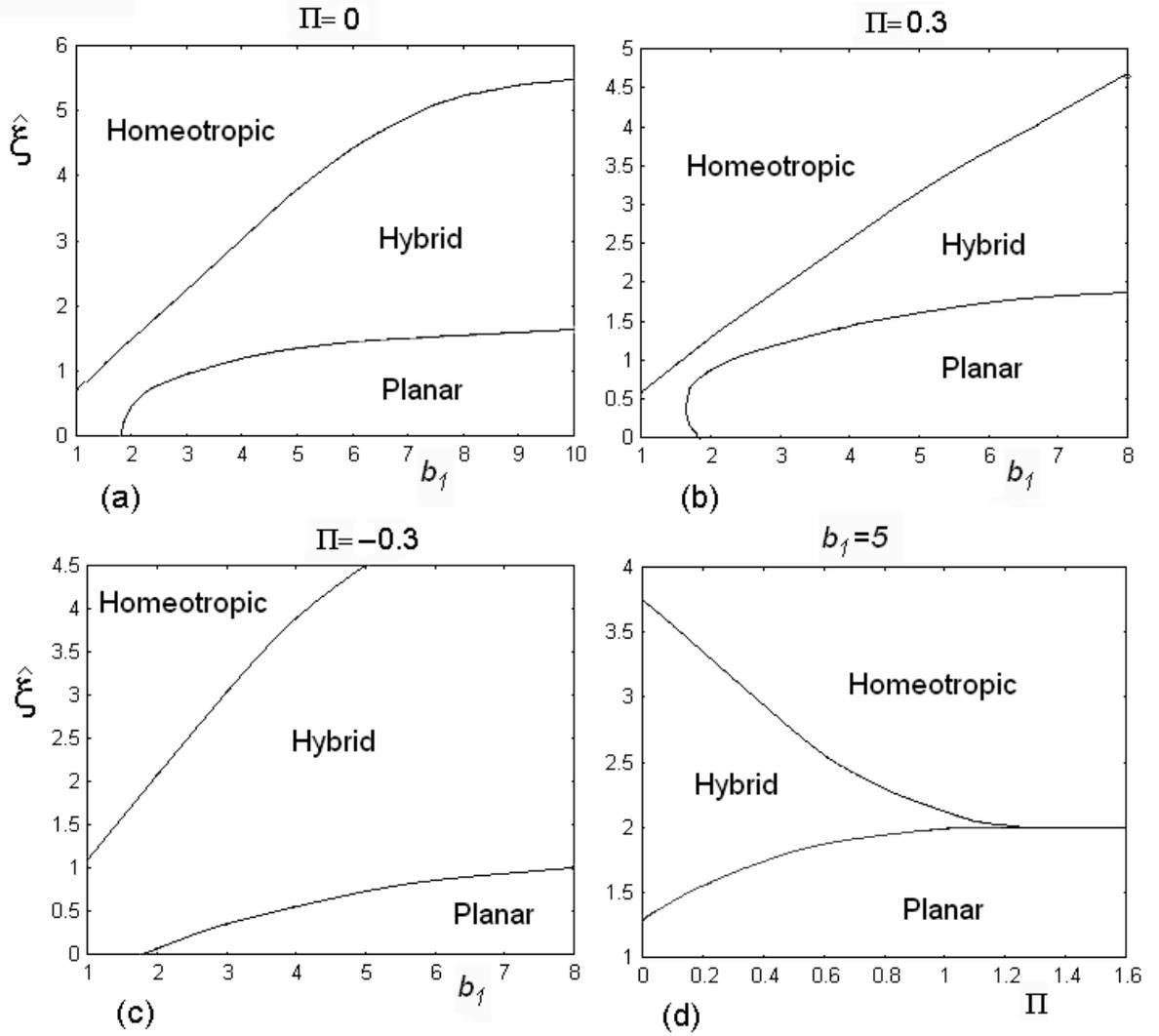

*Fig.5 Phase diagrams of the HAN cell, for a fixed choice of the surface parameter $b_o = 1$. We change the value of parameter $b_1$ and find the value of the thresholds of the electric dimensionless field $\widehat{\xi}$. There are three regions in the diagrams where planar, homeotropic and hybrid alignments are allowed. The phase diagram is depending on the values of flexoelectric parameter $\Pi$. Diagram (d) shows the behaviour of a cell when the flexoelectric parameter $\Pi$ changes. Note that the hybrid configuration disappears when flexoelectric parameter is higher than 1.3.*

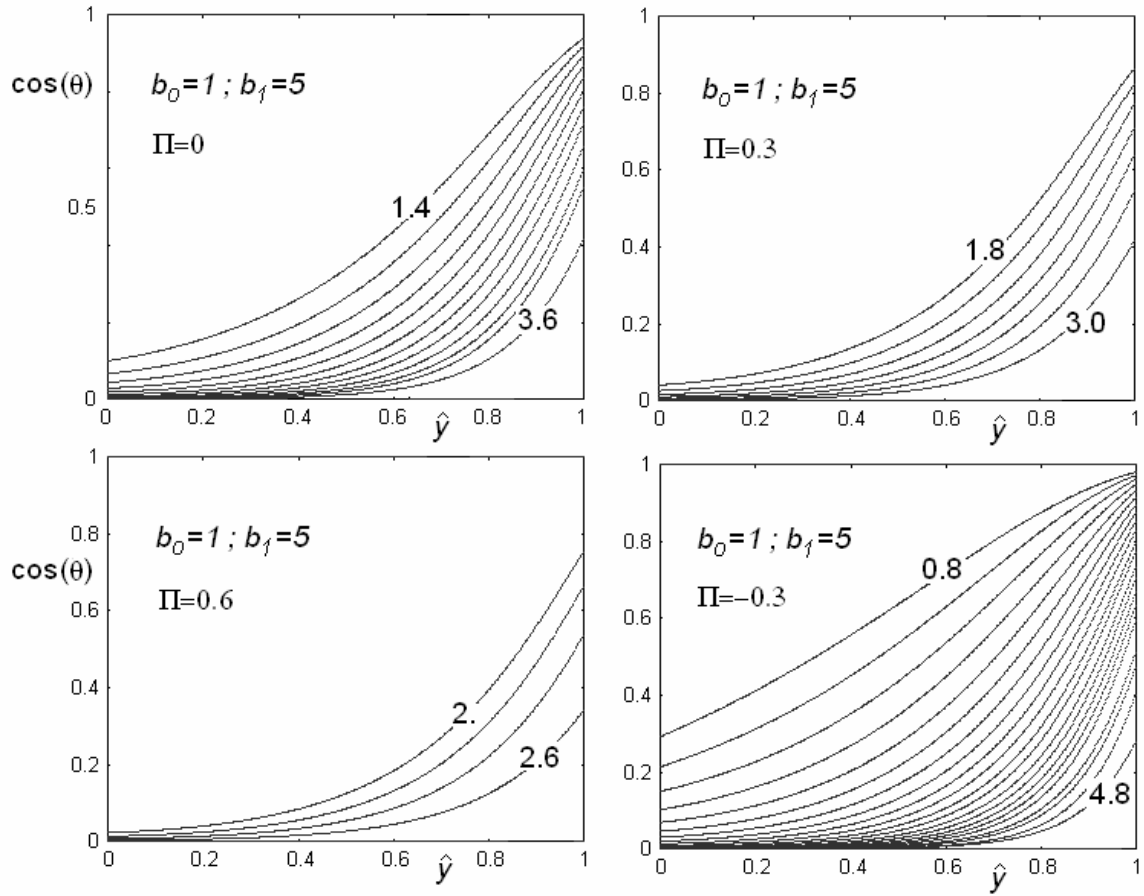

*Fig.6 Behaviour of* $\cos\theta$ *as a function of the reduced cell thickness* $\hat{y} = y/d$ *in the case of positive and negative flexoelectric coefficients, for different values of the dimensionless electric field (some values are reported on the curves). Note that, as the field increases, the role of surface is suppressed and the angle at the planar surface increases. As the electric field is higher that the threshold value, the cell becomes homeotropic and then* $\cos\theta = 0$.

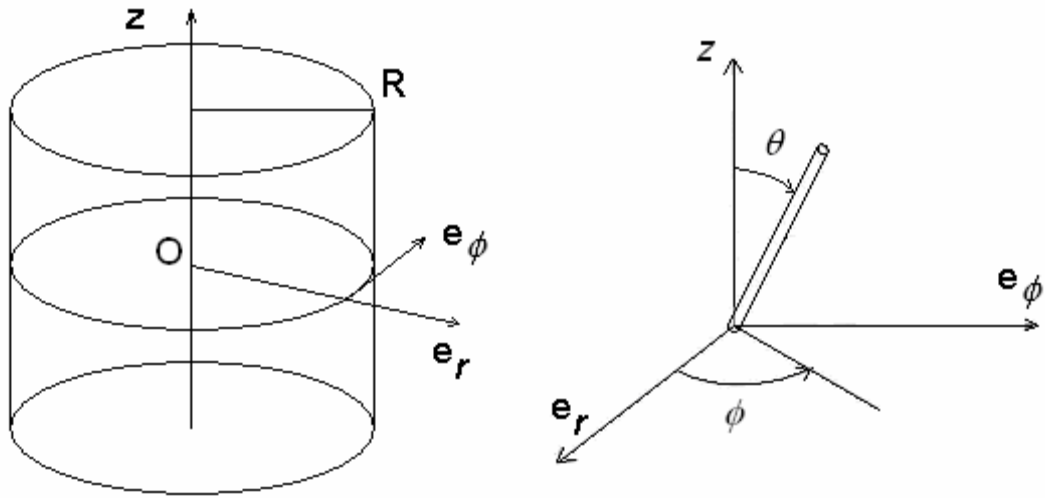

*Fig.7 Cylindrical cell and frame of references on the left and on the right the angles of director chosen for calculations*

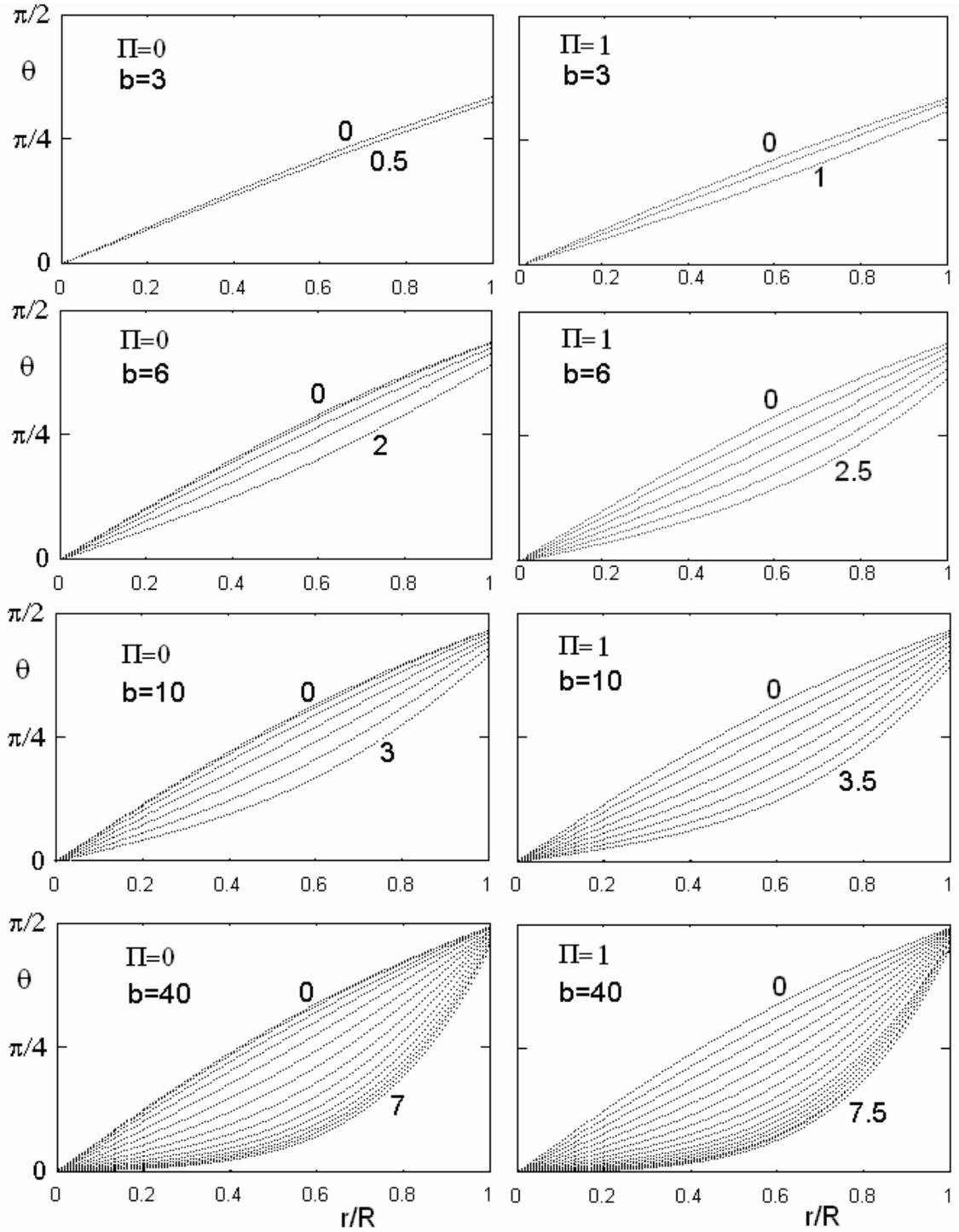

*Fig.8 Behaviour of θ as a function of the reduced radial coordinate r/R in the case of flexoelectric coefficient Π equal to 0 and 1, for different values of the anchoring parameter b and dimensionless electric field (some values are reported on the curves). As the electric field is higher that a threshold value, angle θ goes to zero, that is the director field is parallel to the cylinder axis.*

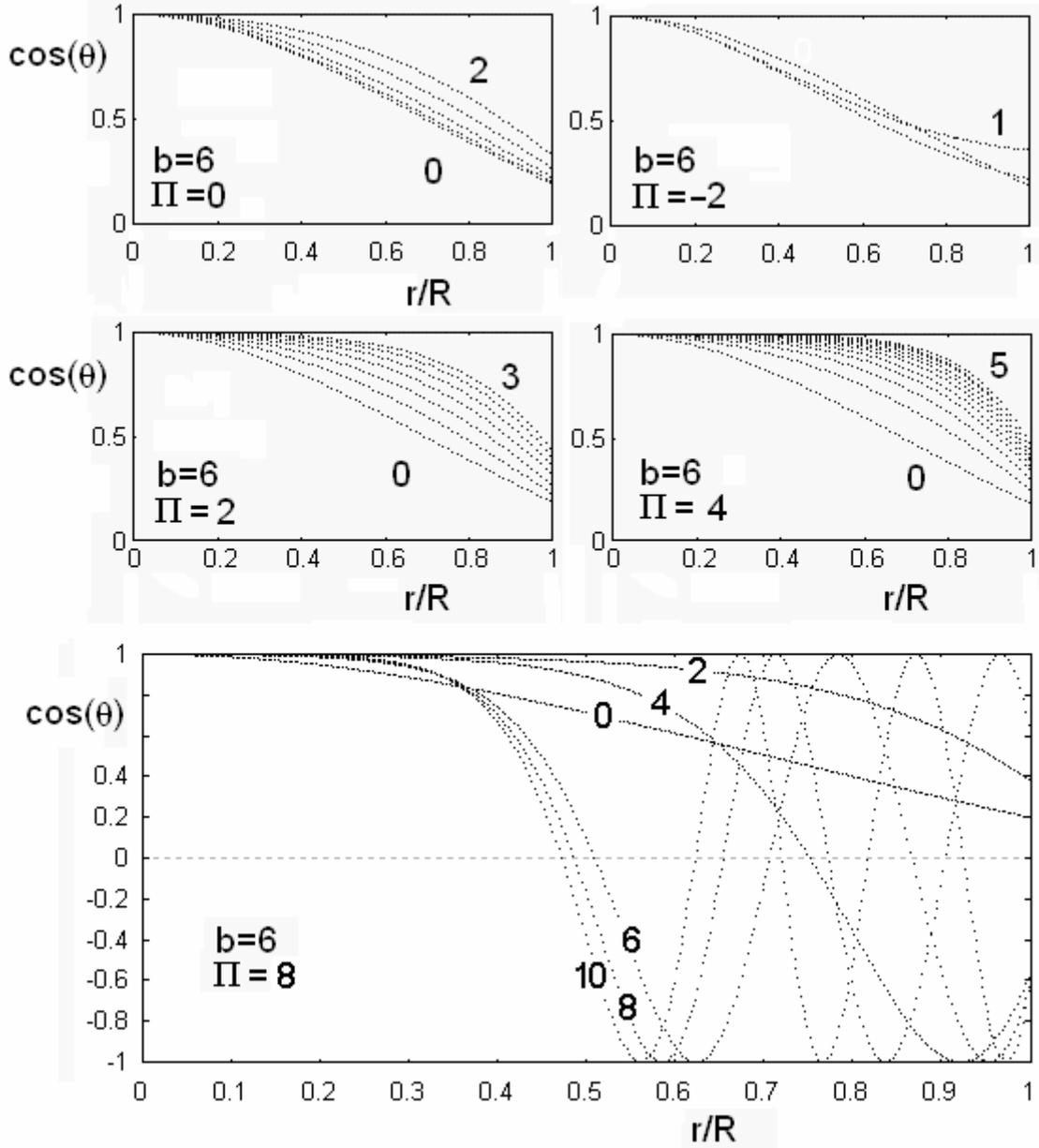

*Fig.9 Behaviour of $\cos\theta$ as a function of the reduced radial coordinate $r/R$, for different values of the flexoelectric coefficient $\Pi$ and of the dimensionless electric field (some values of $\tilde{\xi}$ are reported on the curves). The value of the anchoring strength is fixed in all the figures. Note that a negative value of the flexoelectric parameter is strongly favouring the alignment of director parallel to cylinder axis, and then the threshold electric field is very low. If the flexoelectric parameter is positive and large, the distorted configuration is favoured, and the threshold field, needed for suppressing this configuration, is increased. If $\Pi$ is very large, as in the lower image, $\theta(r)$ is an oscillating function, if the electric field increases. A very large field is required to suppress the distortion and have $\theta = 0$.*

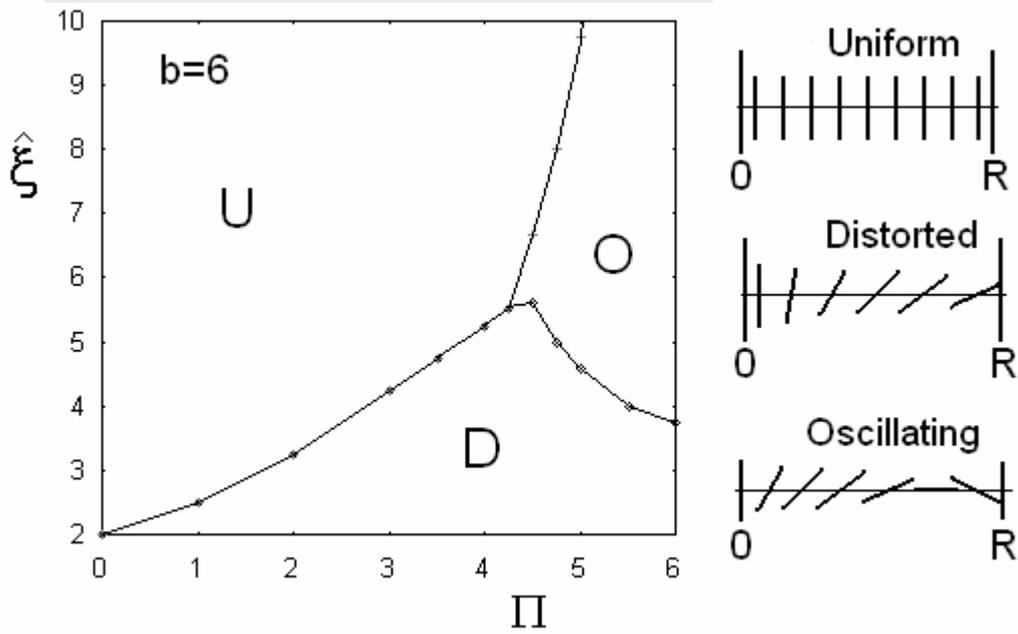

*Fig.10 Phase diagram of the cylindrical confinement, when the anchoring parameter b is fixed and equal to 6. The three regions are denoted by U for the uniform alignment of director parallel to the cylinder axis, D when the director has a deformed configuration, and O if director is oscillating and cosine becomes negative too.*